\documentstyle{elsart}
\begin{document}

\begin{frontmatter}
\title{On the detectability of quantum spacetime foam with 
       gravitational-wave interferometers}
\author{Ronald J. Adler}$^{\mbox{\scriptsize \,a,b,}}\!$\footnote{\scriptsize 
   Corresponding author. E-mail address: {\tt adler@relgyro.stanford.edu}},
\author{Ilya M. Nemenman$^{\mbox{\scriptsize c,a}}$,}
\author{James M. Overduin$^{\mbox{\scriptsize d,a}}$,}
\author{David I. Santiago$^{\mbox{\scriptsize e,a}}$}
\address{$^a$Gravity Probe B, Hansen Experimental Physics Laboratory,
         Stanford University, Stanford, CA 94305, USA}
\address{$^b$Department of Physics, San Francisco State University, 
         San Francisco, CA}
\address{$^c$Department of Physics, Princeton University, Princeton, NJ}
\address{$^d$Department of Physics, University of Waterloo, ON, Canada}
\address{$^e$Department of Physics, Stanford University, Stanford, CA}

\begin{abstract}
We discuss a recent provocative suggestion by Amelino-Camelia and others that
classical spacetime may break down into ``quantum foam'' on distance scales
many orders of magnitude larger than the Planck length, leading to effects
which could be detected using large gravitational wave interferometers.
This suggestion is based on a quantum uncertainty limit obtained by Wigner
using a quantum clock in a gedanken timing experiment.  Wigner's limit,
however, is based on two unrealistic and unneccessary assumptions: 
that the clock is free to move, and that it does not interact with the 
environment. Removing either of these assumptions makes the uncertainty 
limit invalid, and removes the basis for Amelino-Camelia's suggestion.
\end{abstract}

 \endkeyword
\end{frontmatter}

\section{Introduction}

        Amelino-Camelia \cite{AC1,AC2} has made the interesting suggestion 
that the fundamental minimum distance uncertainty between two spatially
separated points may depend on the separation distance and be many orders
of magnitude larger than the Planck length. That is, the classical picture
of spacetime may break down to ``quantum foam'' on a surprisingly large
scale. This would imply that quantum gravity effects could potentially
be probed with current or near-future interferometers designed for use as
gravitational-wave detectors.

        We consider here the arguments put forward by Amelino-Camelia
\cite{AC1,AC2}, as well as related arguments by Ng and van~Dam \cite{NV1,NV2}.
These all rely on a lower limit on distance measurement uncertainty obtained 
long ago by Wigner \cite{W}, who used a quantum clock in a gedanken light travel 
timing experiment. The uncertainty limit is the result of spreading of the wave 
packet of the clock. There are two basic assumptions in the analysis: the 
quantum clock is free, and it does not interact with its environment during 
the course of the (macroscopic) timing experiment.  That is, the clock
evolves according to the unitary operator $U(t,0)=\exp(-iHt)$ with a 
strictly free Hamiltonian; there is no interaction with a potential,
the environment, or anything else.

        We show first that if the clock is quantum mechanical but not free
(if it is bound in a harmonic oscillator potential, for example) then the 
uncertainty limit becomes much smaller than that obtained by Wigner and used
by Amelino-Camelia and Ng and van~Dam.  We then point out that, if the clock
is sufficiently large or complex, it will interact with its environment in
such a way that its wave function decoheres; that is, loses the phase
coherence necessary for superposition into a packet.  In addition, such
interactions (eg, a restraint system) may localize or ``collapse'' the
wave function.  These effects generally happen in a time much less than
that needed for macroscopic distance measurements.  The result is that
the clock wave function does {\em not spread linearly\/} over macroscopic 
times.

        It thus appears that the clock used in the gedanken experiments is
particularly ill-suited to its purpose.  It does not appear to be ``the best
that we can imagine'' \cite{W}.  Indeed, due to decoherence, we expect that
an ideal quantum clock with nontrivial internal construction is not
obtainable, {\em even in principle\/}.

        Finally, we note that an additional gravitational length
uncertainty limit given by Ng and van Dam \cite{NV1,NV2} would violate
everyday experience (by many orders of magnitude) if taken to be in any
way {\em intrinsic\/} to the measurement process.

\section{Review of the quantum uncertainty limit}

        The work of Amelino-Camelia \cite{AC1,AC2} and of Ng and van~Dam
\cite{NV1,NV2} relies on an analysis of distance measurement following
Wigner \cite{W}. The various authors define the spatial distance between two
points as $\ell=ct/2$, where $t$ is the time it takes light to complete a 
round trip between them, and $c$ is of course the velocity of light. At one of
the points they place a system of a clock plus a transmitter-receiver that is
used to send a light signal to a mirror at the other point, and to time
its return. To obtain a quantum limit on the uncertainty in $\ell$ they assume
the clock system behaves as a free quantum object, and calculate the
uncertainty in its position during the transit time of the light. This is a
standard problem discussed in quantum mechanics texts, often referred to as
spreading of the (minimum uncertainty) wave packet (e.g. \cite{Schiff}, p.~64).
It may be solved using a superposition of plane wave solutions of the 
Schr\"odinger equation, or a simple approximation may be obtained using the 
uncertainty principle \cite{NV1,NV2}.

        Briefly the uncertainty principle derivation is as follows. (We
delete factors of order unity throughout as irrelevant to the discussion.)
Denote the initial uncertainty in the clock system position by $\delta\ell_0$
and its mass by $m_c$. By the uncertainty principle this implies an uncertainty 
in its velocity of $\delta v \geq \hbar/m_c\delta\ell_0$. Thus the position 
uncertainty or wave packet width spreads with time, and at $t$ it is 
$\delta\ell(t) \geq \hbar t / m_c \delta\ell_0$. The two uncertainties
$\delta\ell_0$ and $\delta\ell(t)$ combine as independent random variables,
to give a net result
\begin{equation}
\delta\ell_{\scriptscriptstyle Q}^{\, 2} \geq \delta\ell_0^{\, 2} +
    (\hbar t / m_c \delta\ell_0 )^{\, 2}
\label{Eq2.1}
\end{equation}
(Amelino-Camelia and Ng and van~Dam add the two uncertainties linearly,
which gives essentially the same final conclusion.) The net uncertainty is
minimum at $\delta\ell_0^{\, 2} \approx \hbar t/m_c$, so that
\begin{equation}
\delta\ell_{\scriptscriptstyle Q}^{\, 2} \geq \frac{\hbar t}{m_c}
   \approx \frac{\hbar\ell}{m_c c}
\label{Eq2.2}
\end{equation}
This quantum uncertainty limit is taken to be {\em intrinsic\/} to the 
length measurement process.

        We emphasize, as stated in the introduction, that the uncertainty 
limit~(\ref{Eq2.2}) is quite correct for a freely moving quantum clock whose 
wave function evolves according to the Schr\"odinger equation during the 
macroscopic transit time $t$ of the light. We will critically discuss
this limit and its relevance in \S \ref{Sec4}.

\section{Review of the total uncertainty limit} \label{Sec3}

        Amelino-Camelia \cite{AC1,AC2} observes that whatever the nature
of the clock system, it must have a Schwarzschild radius less than its
characteristic size $d$, or the measurement light could not escape the
gravitational field of the clock system and be sent to the mirror; that is,
$Gm_c/c^2 \leq d$. Combined with eq.~(\ref{Eq2.2}) this gives a total 
uncertainty limit of
\begin{equation}
\delta\ell_{\scriptscriptstyle T}^{\, 2} \geq \frac{\ell}{d}
   \frac{G \hbar}{c^3} \approx
   \frac{\ell \ell_{\scriptscriptstyle P}^{\, 2}}{d} \; , \; \; \;
\delta\ell_{\scriptscriptstyle T} \geq \sqrt{
   \frac{\ell \ell_{\scriptscriptstyle P}^{\, 2}}{d} } \; \; \;
   \mbox{(Amelino-Camelia)}
\label{Eq3.1}
\end{equation}
This equation is the basis of ref.~\cite{AC1}, where $d$ is taken to be of
order $\ell_{\scriptscriptstyle P}$. The suggestion is then made that the
uncertainty may be detectable using large interferometers designed for the 
detection of gravitational waves.  Indeed, the uncertainty bound appears
already to be violated experimentally.

        Ng and van~Dam \cite{NV1,NV2} are more specific about the clock and
assume that it contains two mirrors separated by a distance $d$, and that a
pulse of light bounces back and forth between the mirrors. One tick of the
clock takes a time $d/c$, which they take to be a {\em fundamental 
discretization error\/}, $\delta t \geq d/c$. (That is, they assume that the 
light pulse cannot be detected while between the mirrors.) This implies an 
error in the length measurement of $\delta\ell \geq d$.  Finally, the 
assumption is made that the clock system is spherically symmetric and 
larger than its Schwarzschild radius. Then there is a fundamental error 
in the length measurement due to the mass of the clock of roughly
\begin{equation}
\delta\ell_{\scriptscriptstyle G} \geq \frac{G m_c}{c^2}
\label{Eq3.2}
\end{equation}
This gravitational uncertainty limit is taken to be {\em intrinsic\/} to 
the length measurement process. Eq.~(\ref{Eq3.2}) tells us that the fundamental
uncertainty in a distance measurement is approximately equal to the local
deviation of the spacetime metric from Lorentzian. We will critically discuss
this in \S \ref{Sec5}.

        Ng and van~Dam then combine the quantum and gravitational uncertainty 
limit and eliminate the clock system mass by multiplying eqs.~(\ref{Eq2.2})
and (\ref{Eq3.2}), which gives
\begin{equation}
\delta\ell_{\scriptscriptstyle T}^{\, 3} \geq \ell \, \frac{\hbar G}{c^3} =
   \ell \ell_{\scriptscriptstyle P}^{\, 2} \; , \; \; \;
\delta\ell_{\scriptscriptstyle T} \geq
   (\ell\ell_{\scriptscriptstyle P}^{\, 2})^{1/3} \; \; \;
\mbox{(Ng and van~Dam)}
\label{Eq3.3}
\end{equation}
This combined uncertainty limit subsumes the individual quantum and 
gravitational uncertainty limit~(\ref{Eq2.2}) and (\ref{Eq3.2}) in 
their further discussion.

\section{Comments on the quantum uncertainty limit} \label{Sec4}

        Eqs.~(\ref{Eq3.1}) and (\ref{Eq3.3}) indicate that spacetime displays 
quantum foam properties on scales far above the Planck length of $10^{-35}$~m.
For example, a 1~m distance would be fuzzy to about $10^{-18}$~m --- about
the distance presently probed by high energy physics experiments --- according 
to eq.~(\ref{Eq3.1}), and to about $10^{-23}$~m according to eq.~(\ref{Eq3.3}).
Could spacetime really be fuzzy on scales this far beyond the Planck length?
The key factor is clearly the $\ell$ in the quantum relation~(\ref{Eq2.2}).
As we noted, this equation is correct provided the clock system is free and 
evolves according to Schr\"odinger's equation during the position measurement.

        We first comment on the assumption that the clock system is 
{\em free\/}. This is important in that it leads to the factor of $t$ in the 
uncertainty limit~(\ref{Eq2.2}). If we assume the contrary, that the clock is 
bound to other objects in its vicinity, the uncertainty limit does not follow.
As an example, consider a clock bound in a simple harmonic oscillator potential
$V=m_c\omega^2x^2/2$. The width of the ground state wave function for such a 
clock is of order $\delta\ell_{\scriptscriptstyle Q}^2\approx\hbar / m_c\omega$,
and the wave function does {\em not\/} spread with time (e.g. \cite{Schiff},
p.~73). 

In the spirit of the work of the previous authors we may also obtain 
this result heuristically using the uncertainty principle. We think of the 
clock as ``trying'' to settle into a state with zero momentum at the bottom 
of the potential well, but prevented from doing so by the uncertainty principle,
which requires {\em at least\/} that $p\geq\hbar /x$. Then the total energy is
\begin{equation}
E = \frac{p^2}{2m_c} + \frac{m_c\omega^2 x^2}{2}
\label{Eq4.1}
\end{equation}
which has a minimum at $x^2=\hbar/m_c\omega$. This represents the position 
uncertainty of the clock rather than eq.~(\ref{Eq2.2}). In terms of the period
$t_o$ of the oscillator we may thus write the position uncertainty limit as
\begin{equation}
\delta\ell_{\scriptscriptstyle Q}^{\, 2} \geq \frac{\hbar}{m_c\omega}
   = \frac{\hbar t_o}{2\pi m_c}
\label{Eq4.2}
\end{equation}
This has the same form as the uncertainty limit obtained by Wigner, but with 
the time of observation replaced by the period of the oscillator. Whereas the 
time of observation must be macroscopic (as large as desired!) the period of
the oscillator can be made as small as desired in principle --- and quite small
in reality. For example if the clock is taken to be bound like an atom in a
crystal the period would be of order $10^{-12}$~s. It thus appears that the 
assumption that the gedanken clock be free is not necessary and leads to an
unrealistically large distance uncertainty.

        We next comment on the assumption that the clock is truly quantum
mechanical, in the sense that it undergoes unitary evolution according to
the Schr\"odinger equation during the macroscopic time of observation. This
implies that it be sufficiently isolated from its environment that no 
significant interactions occur. If it is not so isolated, its wave function 
will suffer decoherence, which means that the superposition of plane 
waves loses phase coherence and ceases to form a packet.  The decoherence
may be caused by interaction with ambient light, air molecule collisions,
restraint by a tie-down system, or even interaction of the clock with its
{\em own components\/}!  For almost any system of larger than atomic size, 
decoherence occurs in a time much shorter than that required for
macroscopic distance measurement~\cite{Decoh}.  In addition, the interaction
of the clock with its environment (eg, a tie-down system) may be such that
it remains localized.  Loosely speaking, in the language of the Copenhagen 
school, we may think of some interactions as providing position ``measurements''
on a time scale that is less than macroscopic.  A clock that suffers wave
function decoherence or is subject to essentially continuous position 
measurements would not have a linearly increasing position uncertainty, 
and would thus violate the uncertainty limit~(\ref{Eq2.2}).

        In summary it appears that eq.~(\ref{Eq2.2}) is only relevant if
one chooses to consider a freely moving clock undergoing unitary evolution 
according to Schr\"odinger's equation, with no significant environmental 
interactions. Such a clock, if composed of internal mirrors or other parts,
is probably unobtainable, and even if it could be obtained, would be a 
very poor clock.

\section{Further comments on the gravitational uncertainty limit} \label{Sec5}

        Eq.~(\ref{Eq3.2}) of Ng and van~Dam, based on gravity, is interesting
but also presents difficulties. The presence of the measurement clock system
certainly produces a {\em distortion\/} of spacetime, but eq.~(\ref{Eq3.2}) 
tells us that it also produces an {\em uncertainty\/} in spacetime distances
of about the same amount! This is a remarkable statement. Suppose we take it 
seriously as an intrinsic property of spacetime and not just an artifact of 
the model clock. Since the spinning Earth is certainly an excellent clock (the
oldest and most important one we have) we would conclude that objects in the
vicinity of the Earth have a minimum intrinsic position uncertainty of
roughly the Schwarzschild radius of the Earth, which is about 1~cm. This is
manifestly false by many orders of magnitude.

        In general the distances near a massive object of a given
configuration may be determined theoretically (by general relativity) and
measured (by diverse means), and the two agree with each other to impressive 
accuracy, far better than eq.~(\ref{Eq3.2}) would suggest. This is what we mean
when we say that general relativity is well tested \cite{Wil93}.

\section{Comment on the effective hypothetical clock mass} \label{Sec6}

        Amelino-Camelia \cite{AC1,AC2} and Ng and van~Dam \cite{NV1,NV2} use
the uncertainty limits~(\ref{Eq3.1}) and (\ref{Eq3.3}) to discuss the noise in
an interferometer such as the LIGO test model. Both are based on the quantum 
uncertainty limit~(\ref{Eq2.2}). Despite the preceding comments let us suppose 
that eq.~(\ref{Eq2.2}) is correct. Then for small hypothetical clock masses it 
is the operative bound, and we may use noise measurements to place a lower
limit on the effective mass of the hypothetical clock mass. Following 
ref.~\cite{AC1} we use eq.~(\ref{Eq2.2}) and write the variance of the
noise as
\begin{equation}
\delta\ell_{\scriptscriptstyle Q}^{\, 2} = \frac{\hbar t}{2 m_c} =
   \int_{1/t}^{f_{max}} [S(f)]^2 df
\label{Eq6.1}
\end{equation}
where $S$ is the spectral density of the noise due to space time foam in
eq.~(\ref{Eq2.2}). From this it follows that, if the maximum frequency cutoff
$f_{max}$ is reasonably large,
\begin{equation}
S(f) = \sqrt{ \frac{\hbar}{2m_c}} \, \frac{1}{f}
\label{Eq6.2}
\end{equation}
For the LIGO test model the measured noise limit is about
$3 \times 10^{-19}$~m/Hz$^{1/2}$ at 450~Hz \cite{Abr96}, which
places a lower limit on the hypothetical clock mass of about
\begin{equation}
m_c \geq \frac{\hbar}{2 f^2 S^2} \approx 3 \; \mbox{g}
\label{Eq6.3}
\end{equation}
This is a remarkably large mass; it exceeds the mass of the essential
working parts of wristwatches that many of us are wearing at this moment.
As such, it hardly seems plausible as a fundamental property of spacetime.

\section{Summary}

        Our analysis indicates that the quantum uncertainty limit~(\ref{Eq2.2})
is based on assumptions that are neither realistic nor necessary. A quantum
clock bound in a potential well does much better than the postulated limit,
as does a macroscopic clock which interacts continuously with its environment.

        In addition, the gravitational uncertainty limit~(\ref{Eq3.2}) suggested
by Ng and van~Dam appears to imply a fundamental distance uncertainty of about
1~cm near the surface of the Earth, which is contrary to observation.

        Our conclusion is that the uncertainty limits used are artifacts of 
the choice of a particular type of hypothetical clock, and are non-fundamental
in nature. There is thus no reason at present to believe that quantum uncertainty
manifests itself at scales very much larger than the Planck length. 
(For a derivation of the Planck length as the minimum length see
refs.~\cite{RonDavid} and \cite{MTW}.)

        We note in closing that we know of only one way that quantum
spacetime foam might be detectable in the lab in the relatively near future.
Arkani-Hamed, Dimopoulos and Dvali \cite{Dimop} have noted that gravity 
has been probed in the laboratory only down to distances of about 1~cm.
Based partly on considerations of dimensionality, they suggest that gravity may 
operate quite differently at smaller distances, and that the ``effective 
Planck scale'' may consequently be only a little beyond the electroweak 
energy scale now probed in high energy experiments. Experimental tests 
of this idea using table-top sized apparatus will soon be under way.

\end{document}